\documentclass[12pt]{article}

\catcode`\@=11
\def\marginnote#1{}
\newcount\hour
\newcount\minute
\newtoks\amorpm
\hour=\time\divide\hour by60
\minute=\time{\multiply\hour by60 \global\advance\minute
by-\hour}\edef\standardtime{{\ifnum\hour<12
\global\amorpm={am}%
        \else\global\amorpm={pm}\advance\hour by-12 \fi
        \ifnum\hour=0 \hour=12 \fi
        \number\hour:\ifnum\minute<10
0\fi\number\minute\the\amorpm}}
\edef\militarytime{\number\hour:\ifnum\minute<10
0\fi\number\minute}

\def\draftlabel#1{{\@bsphack\if@filesw {\let\thepage\relax
   \xdef\@gtempa{\write\@auxout{\string
      \newlabel{#1}{{\@currentlabel}{\thepage}}}}}\@gtempa
   \if@nobreak \ifvmode\nobreak\fi\fi\fi\@esphack}
        \gdef\@eqnlabel{#1}}
\def\@eqnlabel{}
\def\@vacuum{}
\def\draftmarginnote#1{\marginpar{\raggedright\scriptsize\tt#1}}
\def\draft{\oddsidemargin -.5truein
        \def\@oddfoot{\sl preliminary draft \hfil
        \rm\thepage\hfil\sl\today\quad\militarytime}
        \let\@evenfoot\@oddfoot \overfullrule 3pt
        \let\label=\draftlabel
        \let\marginnote=\draftmarginnote

\def\@eqnnum{(\theequation)\rlap{\kern\marginparsep\tt\@eqnlabel}%
\global\let\@eqnlabel\@vacuum}  }

\def\numberbysection{\@addtoreset{equation}{section}
        \def\theequation{\thesection.\arabic{equation}}}

\def\underline#1{\relax\ifmmode\@@underline#1\else
 $\@@underline{\hbox{#1}}$\relax\fi}

\catcode`@=12
\relax
\numberbysection
\def\beq{\begin{equation}}
\def\eeq{\end{equation}}
\def\bea{\begin{eqnarray}}
\def\eea{\end{eqnarray}}

\def\fin{\end{document}}
\begin{document} 
\begin{titlepage}
\nopagebreak
\begin{flushright}
LPTENS-99/20\\
hep--th/9910235\\
June   1999
\end{flushright}
\vglue 2.5 true cm
\begin{center}
{\large
SOLUTION GENERATING \\
\medskip 
  IN TEN DIMENSIONAL SUPERSYMMETRIC\\ 
\medskip\smallskip  
CLASSICAL YANG--MILLS THEORIES}\\
\vglue 1 true cm
{\itshape \itshape Contribution to the Dubna Memorial Volume\\ In honour of 
Mikhail V.  Saveliev} \\
\vglue 1 true cm
{ Jean--Loup GERVAIS}\\ {\footnotesize Laboratoire de Physique Th\'eorique de
l'\'Ecole Normale Sup\'erieure
\footnote{UMR 8549:  Unit\'e Mixte du Centre National de la Recherche Scientifique, et de 
l'\'Ecole Normale Sup\'erieure. },\\ 24 rue Lhomond, 75231 Paris C\'EDEX
05, ~France.}
\end{center}
\bigskip 
\baselineskip .4 true cm
\noindent
\begin{abstract}
In a recent paper (hep-th/9811108), Saveliev and the author showed that there exits an on-shell
light cone gauge where   the  non-linear part  of the field equations reduces to a (super) version
of Yang's equations which may be solved by methods inspired by the ones previously developed for
self-dual Yang-Mills equations in four dimensions.  Later on (hep-th/9903218), the analogy between
these latter theories  and the present ones was pushed further by writing down a set of super
partial linear differential equations which are the analogues of the Lax pair of Belavin and
Zakharov.  Using this Lax representation, it is shown in the present article that  
solution-generating techniques are at work, which are similar to the ones developed  for four
dimensional self-dual Yang-Mills theories in the late  seventies. 

\end{abstract}

\vfill
\end{titlepage} 
\phantom {
{\large \bf In Memoriam Mikhail V. Saveliev}
}
\vfill \newpage 
\vglue 2 true cm 
\begin{center}
{\large \bf In Memoriam Mikhail V. Saveliev}
\end{center}
\vglue 2 true cm
{\it 
On the morning of September 21st 1998, I  arrive in Durham UK to participate in the 2nd
Annual TMR Conference. I have left Paris rather early in order to deliver my talk at 3 o'clock
concerning the recent progress with Misha. I know that he has not been well and I am much
worried about him; but when Ed. Corrigan greets me and  says ``do you know that Misha died
yesterday?'', I am thunderstruck. I suddenly realise that we will never meet again. How terribly
sad! Our last encounter in person was in Cambridge UK at the beginning of March 1997. He had
accompanied me to the bus station in his usual attentionate and friendly manner, and we were
enthousiastically discussing about future research developments. The evening before he had treated
me to a hearty dinner cooked by him-self, which was the occasion of  a very friendly evening as he
enjoyed so much.  At that time, I felt so sure we would soon meet again! We made plans to do that,
in Brazil  in Paris, or in Tbilissi, but time was always flying too fast. One always assume,
wrongly that the good things will remain for ever for us to reach! Life is unfortunately too short.

 How I miss his frequent email messages! They
were so nice and stimulating. In looking back at them I am impressed by what they brought me day
after day: new ideas, important remarks, key references to the scientific literature,  results of
painstaking calculations, some personal news (too rarely) which gave me a glimpse at the
difficulties of his every day life. In return, and when it came to it, I tried as much as possible
to express my sympathy and support but of course, concerning every day life, Russia is so different
from what I know personally! On top of that, we were fighting  to make progress in our ambitious
program solely by exchanging  ideas over the  net, a notoriously difficult task. I am afraid that he
was under too much pressure, and that this was an important factor in the deterioration of his
health. 

Some parts of his emails bring back so vividely the memory of the time already past:    
On February 1998, Misha wrote from Moscow  

 {\rm Dear Jean--Loup,

Let me answer on yr section ``More difficulties (unfortunately)''.
Definitely you are right, ... Now I understand that you 
was pretty right when said that we should meet personally to put all 
the things in order; unfortunately now it is too late, because I should leave for Brazil already on 25.03.98, Luiz already sent me tickets 
and my grant opens just that time. Nethertheless I am quite optimistic 
that we are on the right track and will succeeded in solving the problem 
in the nearest future....}

He wrote the following from Brazil during Spring 1998: 

{\rm Dear Jean-Loup,

Thanks for yr 3 messages (mail, debut and a latex-file).
..... I'll study yr file in detail
and then be back to you. I am alone here, mainly because
Svetlana Jr has her lessons at the University until June,
then has exams, and since she is, let say a "home child",
we would prefer not to leave her alone, especially since
Moscow is not a quiet small place like Protvino. So it
would be lone for me these three months, but better time
for work. I would say that here it is very nice, Luiz and
other people here, at IFT, are very kind and careful to 
me; meals in the town are nicely in test; I am sure
you would like brazilian food. What do you think about Tbilisi,
will you be able to come? In any case I am writing to my
colleagues in Tbilisi to send you an invitation letter.

With best regards, Misha}

 When
my trip to Brazil was cancelled, he wrote 

{\rm Dear Jean--Loup,

Thank your for your today message.
That's pity that you are not able to come in Sao Paulo; hope
that you'll manage to arrange yr plans to visit Tbilisi, otherwise
I'll try to come in Paris in the Fall at least for a short while
since we have a lot of things to discuss in person, I believe that
see now some remarkable directions in geometry of supermanifolds,
relevant Pluckers, etc. And, as a joke, looking at my directory
GS (Gervais-Saveliev) containing a lot of messages we have exchanged during this
project, I think that we can publish a book of letters with a
title "How many wrong \& right ways might be in science"...}

In retrospect, I am rather struck by the following premonatory part of a message he sent 
me during May 1998: 

{\rm P.S. Last evening I was very busy writing condolences to Russia 
concerning our very eminent physicists David Kirzhnitz from Theory 
Div. of the Lebedev Inst., Polubarinov from Dubna (he was a coauthor
of Victor Ogievetsky), whom both I knew very good for some 30 years,
and Boris Dzelepov, who have passed away two days ago. That's a great  
pity to loose so many colleagues \& friends this year!}

On the other hand,  I
was far from realising that his health had so much deteriorated. He was always blaming his old car
accident. For instance, he wrote en August 28th: 

{\rm Dear Jean-Loup,

Sorry for a long silence. I was in a rather bad 
shape; just after my last message to you I had 
several vascular spasms and swoons, once even 
with a loss of memory, true for a short time. 
Presumably, it is caused by my neck small bell 
problem after the car crash in 1988. Now I am 
slightly better, and am beginning to work, but 
feel necessity to make some medical treatments 
in Moscow. .... }''

Going further back in time, it is a pleasure to remember that I met Misha for the
first time in 1992 when his famous work with Lesnov had already proven to be so important for two
dimensional conformal/integrable systems. We immediately started to collaborate and have done
so, at least on a part time basis,  ever since. His contribution to our research program has been 
invaluable. After a series of papers concerned with two dimensional Toda theories ---where we 
successively discussed black hole solutions, W geometries, and  higher grading  generalisations---
we more recently turned to the explicit classical integration of supersymmetric theories with local
symmetries in more than two dimensions, a very ambitious program which is still far from completion
at the present time. 
     Working with Misha  has been a wonderful experience which terminated so abruptly! I
will always remember our excited and friendly discussions, his kindness and enthousiam, his fantastic
knowledge of the scientific literature!   We, at
\'Ecole Normale,  were lucky enough  to invite him for several extended visits which were extremely
fruitful. Misha and I  met in other places, but altogether much too rarely. 

I will always remember the fun we had in discussing physics;  but I now  regret  that, although we
were very good friends, we seldom took time to socialise outside  research. These
few very warm and friendly encounters are dear to my memory,  especially when his  Svetlana's (as
he used to say)  were present. This happened in particular for a day in my country house during
one of his stay in Paris  and on one evening in his apartment when he was visiting Cambridge UK. 
It is good to remember how happy he was on these occasions, how affectionate and (rightly) proud
he was with his wife and daughter, how friendly and warmly he behaved!   

 M. Saveliev  was great both as a scientist and as a
human being. He was obviously such a good father,  husband, friend!
\bigskip 

\begin{flushright}
Jean-Loup Gervais 
\end{flushright}}
 \vglue 3 true cm 
\section{Introduction} 

\vglue 1 true cm 
In recent  times we turned\cite{GS98} to the classical integration of theories in more than two
dimensions with local extended supersymmetries. Our motivation was twofold. On the one hand this
problem is very important for the recent developments in  duality and M theory. On the other hand, 
the recent advances initiated by Seiberg and Witten indicate that  these theories  are in many ways 
higher dimensional analogues of two dimensional conformal/integrable systems, so that progress may be
expected. Since fall 1997, we have studied  super Yang-Mills theories in ten dimensions. There, it
was shown by Witten\cite{W86} that the field equations are equivalent to  flatness conditions. This
is   a priori  similar to well known basic ones  of Toda theories, albeit no real progress could
be made at that time, since the corresponding Lax type equations  involve an arbitrary light like
vector which plays the role of a spectral parameter.  At first, we reformulated the field
equations  in a way which is similar to    a super version  of    the higher dimensional
generalisations of Toda theories developed by  Razumov and Saveliev\cite{RS97},  where the
Yang-Mills gauge algebra is extended to a super one. This has not yet been published since,
contrary to our initial hope, the two types of theories do not seem to be equivalent. I hope to
return to this problem in a near future.  In the mean time, we found the existence of an on-shell
gauge, in super Yang-Mills where the field equations simplify tremendously and where the first
similarity with self-dual Yang-Mills in four dimensions came out\cite{GS98}. More
recently\cite{G99}, I was able to write down  a set of super partial linear
differential equations whose consistency conditions may be derived from the  SUSY Y-M equations in 
ten dimensions, and which are the analogues of the Lax pair of Belavin and Zakharov\cite{BZ78}.

As is well known, super Yang-Mills theories in ten dimensions just describes a standard non abelian
gauge field coupled with a charged Majorana-Weyl spinor field in the adjoint representation of
the gauge group. The dynamics
is thus specified by the standard action 
\beq
S=\int d^{10} x {\> \rm Tr   }
\left\{
-{1\over 4}Y_{mn}Y^{mn}
+{1\over 2}\bar \phi\left(\Gamma^m \partial_m \phi+\left[X_m,\, \phi\right]_- \right)\right\}, 
\label{action}
\eeq
\beq
Y_{mn}=\partial_mX_n-\partial_nX_m +\left[X_m,\, X_n\right]_-.
\label{F0def} 
\eeq
The notations are as follows\footnote{They are  essentially the same as in ref.\cite{GS98}.}: 
$X_m(\underline x)$ is the vector potential, $\phi(\underline x) $ is the Majorana-Weyl spinor. Both
are matrices in the adjoint representation of the gauge group 
${\bf G}$.  Latin indices
$m=0,\ldots 9$ describe Minkowski components.  Greek indices $\alpha=1,\ldots 16$ denote
chiral spinor components. We will use the superspace formulation with odd coordinates
$\theta^\alpha$. The  super vector potentials, which are valued in the gauge group, are noted  
$A_m\left(\underline x,\underline \theta\right)$, $A_\alpha\left(\underline x,\underline
\theta\right)$. As shown in refs. \cite{W86}, \cite{AFJ88}, we may
remove all the additional fields and uniquely reconstruct the physical fields $X_m$, $\phi$ from
$A_m$ and $A_\alpha$ if we impose the condition $\theta^\alpha A_\alpha=0$ on the latter.

With this condition, it was shown in refs. \cite{W86}, \cite{AFJ88}, that the field equations
derived from the Lagrangian  \ref{action} are equivalent to the flatness conditions 
\beq
{\cal F}_{\alpha \beta=0}, 
\label{flat}
\eeq
where ${\cal F}$ is the supercovariant curvature 
\beq
{\cal F}_{\alpha \beta}=D_\alpha A_\beta+D_\beta A_\alpha+\left[A_\alpha,\, A_\beta\right]+
2\left(\sigma^m\right)_{\alpha\beta}A_m.  
\label{curdef}
\eeq
 $D_\alpha$ denote the superderivatives
\beq
D_\alpha=\partial_\alpha-\left(\sigma^m\right)_{\alpha \beta} 
\theta^\beta {\partial_m}, 
\label{sddef}
\eeq
and we use the Dirac matrices 
\beq
\Gamma^m=\left(\begin{array}{cc}
0_{16\times16}&\left(\left(\sigma^m\right)^{\alpha\beta}\right)\\
\left(\left(\sigma^m\right)_{\alpha\beta}\right)&0_{16\times16}
\end{array}\right),\quad  
\Gamma^{11}= \left(\begin{array}{cc}
1_{16\times16}&0\\0&-1_{16\times16}\end{array}\right).
\label{real1}
\eeq
Throughout the paper, it will be convenient to use the following particular realisation: 
\beq
\left(\left(\sigma^{9}\right)^{\alpha\beta}\right)=
\left(\left(\sigma^{9}\right)_{\alpha\beta}\right)=
\left(\begin{array}{cc}
-1_{8\times 8}&0_{8\times 8}\\
0_{8\times 8}&1_{8\times 8}
\end{array}\right)
\label{real2}
\eeq
\beq
\left(\left(\sigma^{0}\right)^{\alpha\beta}\right)=-
\left(\left(\sigma^{0}\right)_{\alpha\beta}\right)=
\left(\begin{array}{cc}
1_{8\times 8}&0_{8\times 8}\\
0_{8\times 8}&1_{8\times 8}
\end{array}\right)
\label{real3}
\eeq
\beq
\left(\left(\sigma^{i}\right)^{\alpha\beta}\right)=
\left(\left(\sigma^{i}\right)_{\alpha\beta}\right)=\left(\begin{array}{cc}
0&\gamma^i_{\mu,\overline \nu}\\
\left(\gamma^{i\, T}\right)_{\nu,\overline \mu}&0
\end{array}\right),\quad  i=1,\ldots 8. 
\label{real4}
\eeq
The convention for greek letters is as follows: Letters from the beginning of the alphabet run from
1 to 16. Letters from the middle of alphabet run from 1 to 8. In this way,  we shall separate
the two spinor representations of $O(8)$ by rewriting $\alpha_1,\ldots, \alpha_{16} $  as 
$\mu_1,\ldots, \mu_8, \overline \mu_1,\ldots, \overline \mu_8$
   
Using the above explicit realisations on sees that the equations to solve take the form  
\begin{eqnarray}
F_{\mu \nu}\equiv D_\mu A_\nu+D_\nu A_\mu +\left[A_\mu,\,
A_\nu\right]_+&=&2\delta_{\mu\nu}\left(A_0+A_9\right)\label{dynuu}\\
F_{\overline \mu \overline \nu}\equiv  D_{\overline \mu} A_{\overline \nu}+D_{\overline \nu}
A_{\overline \mu} +\left[A_{\overline \mu},\, A_{\overline \nu}\right]_+&=&
2\delta_{{\overline \mu}{\overline \nu}}\left(A_0-A_9\right)\label{dyndd}\\  
F_{ \mu \overline \nu}\equiv   D_{ \mu} A_{\overline \nu}+D_{\overline \nu} A_{ \mu}
+\left[A_{\mu},\, A_{\overline \nu}\right]_+&=&-2\sum_{i=1}^8 A_i\gamma^i_{\mu,\overline
\nu}\label{dynud}
\end{eqnarray}
In my last  paper with M. Saveliev \cite{GS98},  these flatness conditions in superspace were used 
to go to an on-shell light-cone gauge where half of the superfields vanish. After reduction to
$(1+1)$ dimensions, the non-linear part of the equations was transformed into equations for a scalar
superfield which are (super) analogues of the so called Yang equations which were much studied in
connection with solutions of self-dual Yang-Mills  equations in four dimensions. The main
differences between the two type of relations is that derivatives are now replaced by
superderivatives, that there are sixteen equations instead of four, and that the indices are paired
differently. Nevertheless, it was found that these novel features are precisely such that the
equations may be solved by methods very similar to the ones developed in connection with self-dual
Yang-Mills in four dimensions.  The aim of the present paper is to push this analogy much further,
by deriving the analogues of the Lax pair of Belavin Zakharov\cite{BZ78} which was instrumental for
deriving multi-instanton solutions at the end of  the seventies. 

\section{The Lax representation}
For completeness, let us repeat the essential points of ref.\cite{G99}. 
The original  theory is $O(9,1)$ invariant, but the choice of Dirac matrices just summarized is
covariant only under a particular $O(8)$ subgroup. The Lax representation  comes out after
picking up a particular $O(7)$ subgroup of the latter. This done simply by remarking  that we may
choose one $\gamma^i$ to be the unit matrix, in which case the others are antisymmetric and obey
the $O(7)$ Dirac algebra.  This is so, for instance in  the following explicit representation of the
$O(8)$ gamma matrices, where $\gamma^8$ is equal to one, which we will use throughout:   
\begin{eqnarray}
\gamma^1= \tau   _1\otimes \tau   _3\tau   _1\otimes {\bf 1} \quad &\quad 
\gamma^5=\tau   _3\otimes \tau   _3\tau   _1\otimes {\bf 1} \nonumber\\
\gamma^2= {\bf 1}\otimes \tau   _1\otimes \tau   _3\tau   _1 \quad &\quad
\gamma^6= {\bf 1}\otimes \tau   _3\otimes \tau   _3\tau   _1   \nonumber\\
\gamma^3=\tau   _3\tau   _1 \otimes {\bf 1}\otimes \tau   _1 \quad &\quad
\gamma^7= \tau   _3\tau   _1\otimes {\bf 1} \otimes \tau   _3  \nonumber\\
\gamma^4= \tau   _3\tau   _1\otimes \tau   _3\tau   _1\otimes \tau   _3\tau   _1 \quad &\quad
\gamma^8={\bf 1}\otimes  {\bf 1}\otimes {\bf 1}. 
\label{gamdef}
\end{eqnarray}
 With this choice, it follows from equations \ref{dynuu}--\ref{dynud} that 
\beq
F_{\mu \nu}=2\delta_{\mu \nu}\left(A_0+A_g\right),\quad 
F_{\overline \mu \overline \nu}=2\delta_{\overline \mu \overline \nu}\left(A_0-A_g\right),\quad 
F_{\mu \overline \nu}+F_{ \nu \overline \mu}=-4\delta_{ \mu \nu}A_8 . 
\label{symdyn}
\eeq
We have symmetrized the mixed (last) equations so that the right-hand sides only involve 
Kronecker delta's in the spinor indices.  By taking $\gamma^8$  to be the unit
matrix, we have introduced  a  mapping between overlined and non overlined indices.
Accordingly, in the previous equation and hereafter, whenever we write an overlined index and non
overlined one with the same letter  (such as $\mu$ and $\overline \mu$) we mean that they are
numerically equal, so that 
$\gamma^8_{\mu \overline \mu}=1$. Next, in parallel with what was done for self-dual Yang-Mills
in four dimensions, it is convenient to go to complex (super) coordinates. 
Thus we introduce, with $i$ the square root of minus one\footnote{For the new symbols, the group
theoretical meaning of the fermionic indices $\mu$ $\overline \mu$ is lost. We adopt this
convention to avoid clusy notations.},  
$$
G_{\mu \nu}=F_{\mu \nu}-F_{\overline \mu \overline \nu}
+iF_{\overline \mu  \nu}+iF_{\mu \overline \nu} 
$$
$$
G_{\overline \mu \overline \nu}=F_{\mu \nu}-F_{\overline \mu \overline \nu}
-iF_{\overline \mu  \nu}-iF_{\mu \overline \nu}, 
$$
\beq
G_{ \mu \overline \nu}=F_{\mu \nu}+F_{\overline \mu \overline \nu}
+iF_{\overline \mu  \nu}-iF_{\mu \overline \nu},  
\label{Gdef}
\eeq
\beq
\Delta_\mu= D_\mu +iD_{\overline \mu},\quad 
\Delta_{\overline \mu}= D_\mu -iD_{\overline \mu}, 
\label{Deltadef}
\eeq 
\beq
B_\mu= A_\mu +iA_{\overline \mu},\quad  
B_{\overline \mu}= A_\mu -iA_{\overline \mu}.
\label{Bdef}
\eeq
A straightforward computation shows that 
$$
\left[\Delta_\mu,\, \Delta_\nu\right]_+
=4\delta_{\mu \nu}\left(\partial_9-i\partial_8 \right),\quad  
\left[\Delta_{\overline \mu},\, \Delta_{\overline \nu}\right]_+=
4\delta_{\mu\nu}\left(\partial_9+i\partial_8 \right),  
$$
\beq
\left[\Delta_{ \mu},\, \Delta_{\overline \nu}\right]_++
\left[\Delta_{ \nu},\, \Delta_{\overline \mu}\right]_+
=8\delta_{\mu
\nu}\partial_0
\label{anti}
\eeq

Consider, now the system of differential equations
\beq
{\cal D}_\mu\Psi \left(\lambda\right)\equiv 
\left(\Delta_{\mu}+\lambda \Delta_{\overline \mu}+B_{\mu}+\lambda B_{\overline
\mu}\right)\Psi(\lambda)=0,  \mu=1,\ldots, 8. 
\label{BZ}
\eeq
Of course, although we do not write it for simplicity of notations, $\Psi(\lambda)$ is a
superfield function of $\underline x$ and $\underline \theta$. The parameter $\lambda$ is an
arbitrary complex number. The consistency condition of these equations is 
\beq
\left[{\cal D}_\mu,\, {\cal D}_\nu\right]_+\Psi(\lambda)=0. 
\label{cons}
\eeq
This gives 
$$
\left\{4\delta_{\mu \nu}\left(\partial_9-i\partial_8 \right) 
+G_{\mu \nu}\right\}\Psi 
+\lambda\left\{8\delta_{\mu \nu}\partial_0    
+G_{\nu\overline \mu}
+ G_{\overline \mu\nu} \right\}\Psi 
$$
$$
+\lambda^2\left\{ 4\delta_{\mu
\nu}\left(\partial_9+i\partial_8 \right) 
+ G_{\overline \nu\overline \mu}
\right\}\Psi=0. 
$$
Thus we correctly get that, for $\mu\not=\nu$
$$
G_{\mu \nu}=G_{\overline \mu \overline \nu}=
G_{\mu \overline \nu}+G_{ \nu \overline \mu}=0,  
$$
and that $G_{\mu \mu}$ $G_{\overline \mu \overline \mu} $, $
G_{\mu \overline \mu}$ do not depend upon $\mu$. Thus these consistency conditions are equivalent
to the symmetrized dynamical equations \ref{symdyn}.  
\section{Solution generating mechanism}
In this section we discuss a solution generating mechanism analogous to the one developed for
self-dual Yang-Mills in four dimensions in ref\cite{FHP83}. Although this is not absolutly
necessary, we will assume in order to simplify the discussion that there is no dependence upon
$x^i$, for $i=1,\ldots, 8$. Thus, besides the odd fermionic variables,  the superfields only depends
upon
$x_\pm=x^0\pm x^9$.  In ref.\cite{G99}, the hermiticity conditions for superfields were established
with 
$SU(N)$ as the gauge group, assuming to avoid  complications that 
 only look  at solutions such that $\phi^\alpha=0$. For these purely bosonic solutions
$A_{\alpha}$ and $A_{m}$ only involve odd and even powers of $\theta$ respectively. The
hermiticity condition on superfields is that 
\beq
A^\dagger_\alpha=-KA_\alpha K,\quad A^\dagger_m=-KA_m K, 
\label{herm}
\eeq 
 with 
\beq
K=\left(-1\right)^{{\cal R}({\cal R}-1)/2}. 
\label{Kdef}
\eeq  
We will write in general the above hermiticity conditions under the form 
$A^\dagger_\alpha=-\tilde A_\alpha$, $ A^\dagger_m=-\tilde A_m$ and so on.  
Thus 
$$
\tilde B_\mu=-B_{\overline \mu}. 
$$
Assume that we only consider solutions of equation \ref{BZ} which are regular at $\lambda=0$. 
Then it follows that we may write  
\beq
B_\mu=-\left(\Delta_\mu \Lambda\right) \Lambda^{-1},\quad 
\Lambda =\Psi(0). 
\label{Lambdef}
\eeq
Then 
$$
B_{\overline \mu}=-K \Lambda^{\dagger\, -1}\left(\Delta_{\overline \mu }\Lambda^\dagger\right)K=
\left(\Delta_{\overline \mu }\tilde \Lambda^{ -1}\right) \tilde
\Lambda. 
$$
An easy computation then shows that we may change gauge by replacing $\Psi\to \tilde \Lambda
\Psi$, 
in such a way that the Lax equtions become  
\beq
\left(\Delta_\mu +\lambda\Delta_{\overline \mu}-a_\mu\right)\Psi(\lambda)=0
\label{Lax}
\eeq
where $a_\mu=\left(\Delta_\mu g\right)g^{-1}$, and  
\beq
g=\tilde \Lambda\Lambda.  
\label{gdef}
\eeq
  In this gauge, at  $\lambda=0$, we get 
$$
\Delta_\mu\Psi(0)=\left(\Delta_\mu g\right)g^{-1} \Psi_0, 
$$
so that 
$$
\Psi(0)=g. 
$$
Following the discussion of ref.\cite{FHP83} closely, the  input is a solution $\Psi_0$ of
equations
\ref{Lax}, such that
$a^0_{\mu}=\left(\Delta_{\mu}\Psi_0(0)\right)\Psi(0)^{-1}$. We look for a solution 
of equations \ref{Lax} in the following form
\beq
\Psi(\lambda)= \chi(\lambda) \Psi_0(\lambda), 
\label{ansatz1}
\eeq
where 
\beq
\chi(\lambda) ={\bf 1}+\sum_{k=1}^n {R_k\over \lambda-\lambda_k}. 
\label{ansatz2}
\eeq
The quantities  $R_k$ are superfields which are $N\times N$ matrices, and $\lambda_k$ are 
superfields independent from $\lambda$. Hence $\chi$ is meromorphic in the $\lambda $ plane.
Substitute the ansatz \ref{ansatz1} into equations \ref{Lax}. One gets  
\beq
\left(\Delta_{\mu} +\lambda \Delta_{\overline \mu}\right)\chi.\chi^{-1}+
\chi a^0_{\mu}\chi^{-1}.  
=a_{\mu}
\label{chieq}
\eeq
Since $\chi(\lambda)\chi^{-1}(\lambda)=1$, the ansatz \ref{ansatz2} immediately implies that 
\beq
R_k\chi^{-1}(\lambda_k)=0. 
\label{inv}
\eeq
Thus we may write 
\beq
\left(R_k\right)_{ab}=
\sum _{j=1}^{s_k}n_a^{(k, j)}m_b^{(k, j)},\quad 
\left(\chi^{-1}(\lambda_k)\right)_{ab}=
\sum _{\ell=s_k+1}^{N}q_a^{(k, \ell)}p_b^{(k, \ell)}, 
\label{res1}
\eeq 
\beq
m^{(k, j)}. q^{(k, \ell)}\equiv \sum_{a=1}^N m_a^{(k, j)}q_a^{(k,
\ell)}=0, 
\label{res2}
\eeq
where $s_k$ are the dimensions of the subspaces upon which $R_k$ projects. 
We write the above as 
$$
R_k=
\sum _{j=1}^{s_k}n^{(k, j)}\otimes m^{(k, j)},\quad 
\chi^{-1}(\lambda_k)=
\sum _{\ell=s_k+1}^{N}q^{(k, \ell)}\otimes p^{(k, \ell)}. 
$$
The left hand side of equation \ref{chieq} would contain second and first order poles at
$\lambda=\lambda_k$, whereas the right-hand side is analytic.
\paragraph{Absence of double pole}
This leads to the conditions 
\beq
\Delta_\mu \lambda_k+\lambda_k \Delta_{\overline \mu} \lambda_k=0. 
\label{dpoleq}
\eeq
Let us preceed by analogy with the bosonic case. Consider an arbitrary  superfield 
$h(\lambda, x_+,x_-, \theta^1,\ldots \theta^8, \theta^{\overline 1},\ldots \theta^{\overline 8} )$, 
noted $h(\lambda)$ for brevity  satisfying 
\beq
\left(\Delta_\mu +\lambda \Delta_{\overline \mu }
\right)h(\lambda)=0. 
\label{heq}
\eeq
Note that, in terms of $D_\mu$ and $D_{\overline \mu}$ this means that 
$$
\left(D_\mu+iD_{\overline \mu}\right)h+\lambda
\left(D_\mu-iD_{\overline \mu}\right)h, 
$$
so that 
$$
\left(1+\lambda\right)D_\mu h +i\left(1-\lambda\right)D_{\overline
\mu}h=0, 
$$
$$
D_\mu h +i{1-\lambda\over 1+\lambda}D_{\overline \mu}h=0. 
$$
Therefore, the solution of this equation was already derived in ref.\cite{GS98}.  
Equation \ref{dpoleq} is satisfied if $\lambda_k$ are superfields, that is,  
$$
\lambda_k\left(x_+,x_-, \theta^1,\ldots \theta^8, \theta^{\overline 1},\ldots \theta^{\overline 8}
\right), 
$$ 
which are such that 
\beq
h\left (x_+,x_-, \theta^1,\ldots \theta^8, \theta^{\overline 1},\ldots \theta^{\overline 8}
, \lambda_k(x_+,x_-, \theta^1,\ldots \theta^8, \theta^{\overline 1},\ldots \theta^{\overline 8})
\right)\equiv 0. 
\label{solpole}
\eeq
\paragraph{Proof }
Indeed, since we get identically zero, we have
$$
\Delta_\mu h(\lambda_k)=0=\left. \Delta_\mu h\right |_{\lambda {\> \rm fixed  \> }=\lambda_k}+ {\partial
h\over
\partial
\lambda}(\lambda_k) \Delta_\mu
\lambda_k, 
$$
$$
\Delta_{\overline \mu} h(\lambda_k)=0=\left. \Delta_{\overline \mu} h\right |_{\lambda{\> \rm fixed  \>}
=\lambda_k}+{\partial h\over
\partial \lambda}(\lambda_k) \Delta_{\overline \mu}
\lambda_k. 
$$
The result follows from the equation $\Delta_{\mu} h+\lambda \Delta_{\overline \mu} h=0.$
\paragraph{Absence of first order poles}
This leads to the condition  
\beq
\left(\Delta_{\mu}R_k +\lambda \Delta_{\overline \mu}R_k \right).\chi^{-1}(\lambda_k)+
R_k a^0_{\mu}\chi^{-1}(\lambda_k) =0. 
\label{0spoleq}
\eeq
Substitute equation \ref{res1}. One has 
$$
\left(\Delta_{\mu}R_k\right)\chi^{-1}(\lambda_k)= 
\Delta_\mu \left(\sum _{j=1}^{s_k}n^{(k, j)}\otimes m^{(k, j)}\right)
\sum _{\ell=s_k+1}^{N}q^{(k, j)}\otimes p^{(k, j)}
$$
$$=
\sum _{j=1}^{s_k}n^{(k, j)}\otimes \left(  \Delta_\mu m^{(k, j)}\right)
\sum _{\ell=s_k+1}^{N}q^{(k, j)}\otimes p^{(k, j)} 
$$
$$
\left(\Delta_{\overline \mu}R_k\right)\chi^{-1}(\lambda_k)= 
\sum _{j=1}^{s_k}n^{(k, j)}\otimes \left(  \Delta_{\overline \mu} m^{(k, j)}\right)
\sum _{\ell=s_k+1}^{N}q^{(k, j)}\otimes p^{(k, j)}.  
$$
Thus we get
$$
\sum _{j=1}^{s_k}n^{(k, j)}\otimes \left\{  \left(\Delta_\mu+\lambda_k \Delta_{\overline \mu}\right)
m^{(k, j)}\right\}
\sum _{\ell=s_k+1}^{N}q^{(k, \ell)}\otimes p^{(k,  \ell)} 
$$ 
$$
+\left(\sum _{j=1}^{s_k}n^{(k, j)}\otimes m^{(k, j)}\right) a^0_{\mu}
\left(\sum _{\ell=s_k+1}^{N}q^{(k,  \ell)}\otimes p^{(k,  \ell)}\right) =0
$$
Thus we conclude that we have to solve the equations  
 \beq
\left\{  \left(\Delta_\mu+\lambda_k \Delta_{\overline \mu}\right) m^{(k, j)}\right\}
q^{(k, \ell)} 
+ m^{(k, j)} a^0_{\mu} q^{(k, \ell)}=0. 
\label{meq}
\eeq
\paragraph{Solution of these equations}
We observe that $\Psi_0(\lambda_k)$ satisfies
$$
\left(\Delta_{\mu}+\lambda \Delta_{\overline \mu}-a^0_{\mu}\right)\Psi_0(\lambda)=0.
$$
Thus we get 
$$
-\left(\Delta_{\mu}+\lambda \Delta_{\overline \mu}\right) \Psi^{-1}_0(\lambda) 
-\Psi^{-1}_0(\lambda)a^0_{\mu}=0, 
$$
and, setting $\lambda=\lambda_k$, 
$$
-\left(\Delta_{\mu}+\lambda_k \Delta_{\overline \mu}\right) \Psi^{-1}_0(\lambda_k) 
-\Psi^{-1}_0(\lambda_k)a^0_{\mu}=0. 
$$
Thus equations \ref{meq} are solved by 
\beq
m_a^{(k,j)}=M_b^{(k,j)}\left(\Psi^{-1}_0\right)_{ba}(\lambda_k), 
\label{mres}
\eeq
where $M_b^{(k,j)}$ is a solution of the equation 
\beq
\left(\Delta_{\mu}+\lambda_k \Delta_{\overline \mu}\right)M_b^{(k,j)}=0.  
\label{Meq}
\eeq
\paragraph{Hermiticity}
According to equation \ref{gdef} 
\beq
g=\tilde g. 
\label{herm2} 
\eeq
A straightforward computation gives 
$$
\left(\Delta_{\mu} +\lambda \Delta_{\overline  \mu}-
\Delta_{ \mu} g. g^{ -1} \right)g\tilde \Psi^{
-1}(1/\lambda^*)  =0. 
$$  
We see that $g \tilde \Psi^{-1}(1/\lambda^*)$ and $\Psi(\lambda)$ satisfy the same
equation.
Thus we may assume that  $g \tilde \Psi^{-1}(1/\lambda^*)=\Psi(\lambda)$.
Let us assume that the simple solution satisfies the same condition:
$g_0\tilde \Psi_0^{ -1}(1/\lambda^*)=\Psi_0(\lambda)$. Using the ansatz
\ref{ansatz2}, we see that 
$$
g\tilde \chi^{ -1}(1/\lambda^*)\, \tilde \Psi_0^{ -1}(1/\lambda)=\chi(\lambda)
\Psi_0(\lambda). 
$$
Thus  we get 
$$
g\tilde \chi^{ -1}(1/\lambda^*)\, g^{-1}_0\Psi_0(\lambda)=\chi(\lambda)
\Psi_0(\lambda). 
$$
Therefore, we arrive at the condition
\beq
g\tilde \chi^{
 -1}(1/\lambda^*) \, g^{-1}_0=\chi(\lambda).  
\label{hermchi}
\eeq
Writing  equivalently
\beq
g  =\chi(\lambda)g_0\tilde \chi(1/\lambda^*), 
\label{hermg}
\eeq
we see that, at the poles  $\chi(1/\lambda^*_k)=0$. 
Thus $\chi^{-1}(\lambda)$ has poles at 
$\lambda=1/\lambda^*_k$. Take the residue of equation \ref{hermg} at $\lambda=\lambda_k^{* -1}$. 
According to equation \ref{ansatz2},
$$
\chi^\dagger(1/\lambda^*_k) ={\bf 1}+\sum_{\ell=1}^n {R^\dagger_\ell\over
\lambda_k^{* -1}-\lambda_\ell}, 
$$
which gives, according to equation \ref{res1}, 
$$
\left(\chi(1/\lambda^*_k) \right)_{ab}=\delta_{ab}+\sum_{\ell=1}^n 
\sum_{j=1}^{s_\ell}{n_a^{(\ell, j)} m_b^{(\ell, j)}\over
\lambda_k^{* -1}-\lambda_\ell}. 
$$
The residue must vanish. This gives 
$$
0=\chi(1/\lambda^*_k)  g_0 R_k^\dagger=
\left(\delta_{ad}+\sum_{\ell=1}^n 
\sum_{j=1}^{s_\ell}{n_a^{(\ell, j)} m_d^{(\ell, j)}\over
\lambda_k^{*
-1}-\lambda_\ell}\right)\left(g_0\right)_{dc}\left(R^*_k\right)_{bc}. 
$$
Thus we find that 
$$
\left(\delta_{ad}+\sum_{\ell=1}^n 
\sum_{j=1}^{s_\ell}{n_a^{(\ell, j)} m_d^{(\ell, j)}\over
\lambda_k^{* -1}-\lambda_\ell}\right)\left(g_0\right)_{dc} m_c^{*(k,
j_k)}=0,  
$$
that is, 
$$
{1\over \lambda^*_k}\left(g_0\right)_{ac}\bar m_c^{(k, j_k)}+ \sum_{\ell=1}^n 
\sum_{j_\ell=1}^{s_\ell}{n_a^{(\ell, j_\ell)} m_d^{(\ell, j_\ell)}\over
1-\lambda^*_k\lambda_\ell}\left(g_0\right)_{dc}  m_c^{*(k, j_k)}=0. 
$$ 
Define 
\beq
\Gamma^{(\ell, j_\ell, k, j_k)}\equiv  
{ m_d^{(\ell, j_\ell)} \left(g_0\right)_{dc} m_c^{*(k, j_k)}\over
1-\lambda^*_k\lambda_\ell}. 
\label{Gamdef}
\eeq
We have now 
$$
{1\over \lambda^*_k}\left(g_0\right)_{ac} m_c^{*(k, j_k)}+ \sum_{\ell=1}^n 
\sum_{j_\ell=1}^{s_\ell}n_a^{(\ell, j_\ell)} \Gamma^{(\ell, j_\ell, k, j_k)}=0. 
$$
Thus, 
\beq
n_a^{(k, j_k )} =- \sum_{\ell=1}^n \sum_{j_\ell=1}^{s_\ell}
{1\over \lambda^*_\ell}\left(g_0\right)_{ac} m_c^{*(\ell, j_\ell)} 
\Gamma^{-1\,( \ell, j_\ell, k, j_k)}. 
\label{nres}
\eeq
Substitute finally into equation \ref{ansatz2}, which gives
$$
g=\Psi(0)=\chi(0)g_0=\left({\bf 1}-\sum_k {R_k\over
\lambda_k}\right)g_0; 
$$
that is, according to equation \ref{res1},  
$$
\left(g\right)_{ab}=\left(g_0\right)_{ab}-
\sum_k \sum _{j_k=1}^{s_k}{
n_a^{(k, j_k)}
m_c^{(k, j_k)}\over \lambda_k}\left(g_0\right)_{cb}. 
$$
Substitute equation \ref{nres}. One gets 
$$
\left(g\right)_{ab}=\left(g_0\right)_{ab}+
\sum_k \sum _{j_k=1}^{s_k}
\sum_{\ell=1}^n \sum_{j_\ell=1}^{s_\ell}
{1\over \lambda_k \lambda^*_\ell}\left( g_0\right)_{ad} m_d^{* (\ell, j_\ell)} 
\Gamma^{-1\,( \ell, j_\ell, k, j_k)}
m_c^{(k, j_k)}\left(g_0\right)_{cb}. 
$$
Define 
\beq
N_b^{(k, j_k)}\equiv  m_c^{(k, j_k)}\left(g_0\right)_{cb}. 
\label{Ndef}
\eeq
Since $\tilde g_0=g_0$, we have $\left(g_0\right)_{ab}=\left( \tilde g_0\right)_{ba}$. Thus 
$$
 \tilde N_b^{(k, j_k)}\equiv   \tilde m_c^{(k,
 j_k)}\left(g_0\right)_{bc}. 
$$
Thus we finally arrive at the following expression for $g$ 
\beq
\left(g\right)_{ab}=\left(g_0\right)_{ab}+
\sum_{k=1}^n \sum _{j_k=1}^{s_k} 
\sum_{\ell=1}^n \sum_{j_\ell=1}^{s_\ell}
{1\over \lambda_k \lambda^*_\ell}
\tilde N_a^{(\ell, j_\ell)}
\Gamma^{-1\,( \ell, j_\ell, k, j_k)}
 N_b^{(k, j_k)}. 
\label{Sol1}
\eeq
Concerning $\Psi$, we obtain   
$$
\Psi(\lambda)=\chi(\lambda)\psi_0\left(\lambda\right)=\left({\bf 1}+\sum_k {R_k\over
\lambda-\lambda_k}\right)\psi_0(\lambda), 
$$
that is, according to equation \ref{res1},  
$$
\Psi_{ab}(\lambda)=\chi(\lambda)_{ac}\psi_{0\, cb}\left(\lambda\right)
=\left(\delta_{ac}+\sum_k \sum_{j_k=1}^{s_k}
{n_a^{(k, j_k)} m_c^{(k, j_k)}\over \lambda-\lambda_k}\right)\psi_{0\,
cb}(\lambda). 
$$ 
Substitute equation \ref{nres}. One gets
$$
\Psi_{ab}(\lambda)
=
$$
\beq
\left(\delta_{ac}-\sum_{k=1}^n \sum_{j_k=1}^{s_k}
\sum_{\ell=1}^n \sum_{j_1=1}^{s_1}{1\over \lambda_1(\lambda-\lambda_k)}
\left( g^{-1}_0\right)_{ad} \tilde m_d^{(\ell, j_1)} 
\Gamma^{-1\,( \ell, j_1, k, j_k)}
 m_c^{(k, j_k)}\right)\psi_{0\, cb}(\lambda). 
\label{Psisol}
\eeq
\paragraph{Question of determinent}
Since the gauge group is taken to be $SU(N)$ we have to impose that $\det g=1$. Let us try to
impose this condition, since is not satisfied yet. For this purpose  we want to compute the
determinent of equation
\ref{Sol1}. Consider the case of only one pole first. One has, in that
case, 
$$
\Psi_{ab}(\lambda)
=\left(\delta_{ac}- \sum_{j_1=1}^{s_1}
 \sum_{j'_1=1}^{s_1}{1\over \lambda_1(\lambda-\lambda_1)}
\left( g_0\right)_{ad} m_d^{*(1, j'_1)} 
\Gamma^{-1\,( 1, j'_1, 1, j_1)}
 m_c^{(1, j_1)}\right)\Psi_{0\, cb}(\lambda). 
$$
Consider 
$$
P_{ac}\equiv {1\over 1- \lambda_1\lambda^*_1} \sum_{j_1=1}^{s_1}
 \sum_{j'_1=1}^{s_1} \left( g_0\right)_{ad} m_d^{*(1, j'_1)} 
\Gamma^{-1\,( 1, j'_1, 1, j_1)}
 m_c^{(1, j_1)}. 
$$
It is easily verified that it is a projector such that  
$$
{\> \rm Tr  \> }P\equiv \sum_a P_{aa}=
{\> \rm Tr  \> } \left(\Gamma^{-1}\Gamma\right)=s_1. 
$$
We may write 
$$
\Psi(\lambda)=\left({\bf 1}-{1-\lambda_1\lambda_1^*
\over \lambda^*_1(\lambda-\lambda_1)}P \right)
\psi_{0}(\lambda). 
$$
After diagonalising, it is trivial that, if there is only one pole, 
\beq
{\> \rm det  \> }\Psi\left(\lambda\right)= \left({\lambda -\lambda^{*-1}_1
\over \lambda-\lambda_1}\right)^{s_1}{\> \rm det  \> }\Psi_0. 
\label{det1}
\eeq  
If there are several poles, we may include them in succession by writing 
$$
{\bf 1}+\sum_k {R_k\over \lambda-\lambda_k}=\prod_k 
\left({\bf 1}+ {\tilde R_k\over \lambda-\lambda_k}\right). 
$$
Then we get
\beq
{\> \rm det  \> }\Psi\left(\lambda\right)= \prod_k \left({\lambda -\lambda^{*-1}_k
\over \lambda-\lambda_k}\right)^{s_k}{\> \rm det  \> }\Psi_0, 
\label{detn1}
\eeq  
\beq
{\> \rm det  \> }{\bf g}= \prod_k |\lambda_k|^{-2s_k}{\> \rm det  \>
}\Psi_0. 
\label{detn2}
\eeq 
Thus the determinent is not one. Fortunately, if we have 
$\left(\Delta_{\mu}+\lambda \Delta_{\overline \mu}-a_{\mu}\right)\Psi=0$, we have 
$$
\left(\Delta_{\mu}+\lambda \Delta_{\overline \mu}\right)\ln {\> \rm det  \> } \Psi=
\sum_{ab} \left(\Psi^{-1}\right)_{ab}
\left(\Delta_{\mu}+\lambda \Delta_{\overline \mu}\right) \left(\Psi\right)_{ba}
$$
$$
=\sum_{ab} \left(\Psi^{-1}\right)_{ab}
\left(a_{\mu}\Psi\right)_{ba}=\sum_{a}\left(a_{\mu}\right)_{aa}=
-\sum_{ab}\left(g\right)_{ab}\Delta_\mu \left(g_{-1}\right)_{ba}=
 -\Delta_\mu \ln { \rm det  \> }g.  
$$
Thus $\widetilde \Psi=\left(\det \Psi\right)^\nu \Psi$ also solves equations \ref{Lax}, but now
with $g \left(\det g\right)^{-\nu}$. Accordingly, we define the physical $g$ as 
\beq
g_p=g\prod_{k=1}^n |\lambda_k|^{2s_k}. 
\label{gphys}
\eeq
This terminates the derivation of the multipole solution. 
\section{Outlook}
We have verified that the solution generating method developed for the self-dual Yang-Mills in four
dimensions may be straighforwardly applied to our case.  This is yet another indication that  the
present system of  symmetrised
 equations \ref{symdyn} is indeed completely and explicitly integrable. 

Concerning the full Yang-Mills equations or equivalently the unsymmetrised equations
\ref{dynuu}--\ref{dynud}, any solution is also a solution of the symmetrised equations
\ref{symdyn}. Thus we should be able to derive solutions of the latter which are general enough so
that we may impose that they be solutions of the former. This problem is currently under
investigation. 
\bigskip 

\noindent { \bfseries\Large Acknowledgements: }\\
It is a pleasure to acknowledge stimulating discussions with P. Forg\'acs and H.~Samtleben.

\end{document}